\documentclass[12pt]{article}
\usepackage{epsfig}
\usepackage{citesort}
\date{\today}
 \normalsize

\newcommand{\bmat}{\left(\begin{array}}
\newcommand{\emat}{\end{array}\right)}
\newcommand{\be}{\begin{equation}}
\newcommand{\ee}{\end{equation}}
\newcommand{\bea}{\begin{eqnarray}}
\newcommand{\eea}{\end{eqnarray}}
\def\ie{{\it i.e.}}
\def    \be            {\begin{equation}}
\def    \ee            {\end{equation}}
\def    \bea           {\begin{eqnarray}}
\def    \eea           {\end{eqnarray}}
\def\ie{{\it i.e.}}
\def\lsim{\raise0.3ex\hbox{$\;<$\kern-0.75em\raise-1.1ex\hbox{$\sim\;$}}}
\def\gsim{\raise0.3ex\hbox{$\;>$\kern-0.75em\raise-1.1ex\hbox{$\sim\;$}}}

\begin{document}
\renewcommand{\thefootnote}{\fnsymbol{footnote}}
\rightline{\today} \vspace{.5cm} {\large
\begin{center}
{\large \bf Neutrino masses, mixing and leptogenesis in TeV scale
$B-L$ extension of the standard model}
\end{center}}
\vspace{.2cm}
\begin{center}
M . Abbas  and S. Khalil\\
\vspace{.3cm}
\emph{Center for Theoretical Physics at the
British University in Egypt, Sherouk City, Cairo 11837, Egypt.\\
Faculty of Science, Ain Shams University, Cairo 11566, Egypt.}
\end{center}
%
\hrule \vskip 0.3cm
\begin{abstract}
We address the issue of the neutrino masses and mixing in TeV
scale $B-L$ extension of the Standard Model. We show that if Dirac
neutrino masses are of order $10^{-4}$ Gev, then the measured
neutrino masses are correctly obtained. We propose a mass relation
between quarks and leptons that may account for such small Dirac
neutrino masses. We analyze the leptogenesis in this type of
models and provide analytical expressions for the new
contributions due to the predicted extra Higgs and extra neutral
gauge boson. We find that thermal leptogenesis, with a resonant
enhancement due to nearly degenerate right-handed neutrinos, can
yield sufficient baryon asymmetry. Finally, we comment on a
possible scheme for non-thermal leptogenesis, which is due to the
decay of extra Higgs into right-handed neutrino.
\end{abstract}
\vskip 0.3cm \hrule \vskip 0.5cm
\section{{\large \bf Introduction}}
The Standard Model (SM) of electroweak and strong interactions has
had a tremendous success when confronted with experiment. However,
non-vanishing neutrino masses provides the first confirmed hint
towards physics beyond the SM. The evidence of very light neutrino
masses is now well established by measuring neutrino oscillations
in solar and atmospheric neutrinos. It was shown that the minimal
extension of the SM gauge group by an extra $U(1)$ gauge symmetry
has all the necessary requirements to generate the observed
neutrino masses \cite{Mohapatra:1980qe}. In particular, this type
of models has three SM singlet fermions that arise as a result of
the anomaly cancellation conditions. These particles account for
right-handed neutrinos and give a natural explanation for the
seesaw mechanism.

Recently, a TeV scale $B-L$ symmetry breaking, based on the gauge
group $G_{B-L}\equiv SU(3)_{C}\times SU(2)_{L}\times
U(1)_{Y}\times U(1)_{B-L}$, has been studied \cite{Khalil:2006yi}.
It was emphasized that this model can account for the current
experimental results of the light neutrino masses and their large
mixing. In addition, it predicts an extra SM singlet scalar (extra
Higgs) and an extra neutral gauge boson corresponding to $B-L$
gauge symmetry \cite{Khalil:2006yi}. These new particles may have
significant impact on the SM phenomenology, hence lead to
interesting signatures at Large Hadron Collider (LHC)
\cite{Emam:2007dy}.

On the other hand, the observed baryon asymmetry in the universe
provides another indication for physics beyond the SM since it has
been established that the strength of the CP violation in the SM
is not sufficient to generate this asymmetry \cite{bau}. The CP
violating decay of the right handed neutrinos may generate
leptogenesis which is the most attractive mechanism to interpret
this baryon asymmetry. Since the evidence of non-vanishing
neutrino masses, there has been a growing interest concerning
leptogenesis and it becomes a competent to the other baryogenesis
mechanisms \cite{Chen:2007fv}. It is interesting to note that a
possible solution to the two major problems, which represent the
most serious evidences for new physics, can be naturally emerged
in such simple extension of the SM.

The aim of this paper is to explore the issue of the neutrino
masses, mixing, and leptogenesis in this type of low scale $B-L$
extension of the SM. We show that a low scale $B-L$ symmetry
braking assists to find a mass relation between quarks and
leptons, which may arise from a flavor symmetry. In this case, one
can determine the unknown Dirac neutrino mass matrix, which is an
essential for evaluating the lepton asymmetry. A detailed
investigation for the leptogenesis in this type of models will be
provided. We compute the new one-loop contributions to the decay
of the lightest right handed neutrino due to the new extra Higgs
and extra gauge boson.

Since these contributions do not include any strong CP violating
phase, they do not interfere with the tree level contribution and
hence they have no direct impact on the CP asymmetry. However, due
to the fact that their size can be of the same order as the tree
level one, their interference with the SM loop contributions can
be relevant and enhance significantly the asymmetry. If the $B-L$
effect is negligible respect to the tree level, then the thermal
leptogenesis is viable only if the right-handed neutrinos are
nearly degenerate in mass so that the lepton asymmetry is
resonantly enhanced. We show that $B-L$ contribution of order the
tree level contribution can relax this degeneracy constraint. The
non-thermal leptogenes is another interesting possibility for
enhancing the lepton asymmetry. In this case, the lightest right
handed neutrino can be decay product of heavier particle like, for
instance, the $B-L$ extra Higgs or extra gauge boson. We show that
this is a feasible scenario and is likely to take place in our
model. However the out of equilibrium condition impose strong
constrain on the mass of the heavy particle and on its coupling
with right-handed neutrino. Therefore, the baryon asymmetry in
non-thermal leptogenesis can be enhanced by two order of magnitude
at most than that of thermal scenario.

The paper is organized as follows. In section 2 we briefly discuss
the $B-L$ symmetry breaking and explore the possible constraints
on the corresponding scale. Section 3 is devoted for neutrino
masses and mixing in our TeV scale $B-L$ extension of the SM. We
show that light neutrino masses can be generated through the
seesaw mechanism if the Dirac neutrino masses are of order
$10^{-4}$ GeV. This range of Dirac neutrino mass is consistent
with possible relations may be obtained between the observed quark
and lepton masses. In section 4 we investigate the lepton
asymmetry due to the decay of lightest right-handed neutrino and
analyze both scenarios of thermal and non-thermal leptogenesis.
Finally we give our concluding remarks in section 5.

%
\section{{\large{\bf $B-L$ symmetry breaking}}}
We start our analysis by considering different scenarios of $B-L$
symmetry breaking. The relevant part for the Lagrangian of the
leptonic sector in the minimal extension of the SM $SU(3)_C \times
SU(2)_L \times U(1)_Y \times U(1)_{B-L}$ is given by%
\bea%
{\cal L}_{B-L}&=&-\frac{1}{4} C_{\mu\nu}C^{\mu\nu} + i~ \bar{l}
D_{\mu} \gamma^{\mu} l + i~ \bar{e}_R D_{\mu} \gamma^{\mu} e_R +
i~ \bar{\nu}_R D_{\mu} \gamma^{\mu} \nu_R + (D^{\mu}
\phi)^\dagger(D_{\mu} \phi) \nonumber\\
&+& (D^{\mu} \chi)^\dagger(D_{\mu} \chi)- V(\phi, \chi)-
\Big(\lambda_e \bar{l} \phi e_R + \lambda_{\nu} \bar{l}
\tilde{\phi} {\nu}_R +
\frac{1}{2} \lambda_{\nu_R} \bar{\nu_R^c} \chi \nu_R+\lambda_\nu \bar{\nu_R^c}\tilde{\phi} l^c+ h.c.\Big),~~~~%
\label{lagrangian}
\eea %
where $C_{\mu\nu} = \partial_{\mu} C_{\nu} - \partial_{\nu}
C_{\mu}$ is the field strength of the $U(1)_{B-L}$. The covariant
derivative $D_{\mu}$ is generalized by adding the term $i g^{''}
Y_{B-L} C_{\mu}$, where $g^{''}$ is the $U(1)_{B-L}$ gauge
coupling constant and $Y_{B-L}$ is the $B-L$ quantum numbers of
involved particles. The $Y_{B-L}$ for leptons and Higgs are given
by: $Y_{B-L}(l) =-1$, $Y_{B-L}(e_R) =-1$, $Y_{B-L}(\nu_R) =-1$,
$Y_{B-L}(\phi) =0$ and $Y_{B-L}(\chi) =2$. In
Eq.(\ref{lagrangian}), $\lambda_{e}$, $\lambda_{\nu}$ and
$\lambda_{\nu_R}$ refer to $3\times 3$ Yakawa matrices.

In order to analyze the $B-L$ and electroweak symmetry breaking,
we consider the most general Higgs potential invariant under
these symmetries, which is given by%
\bea%
V(\phi,\chi)&=&m_1^2 \phi^\dagger \phi+m_2^2 \chi^\dagger
\chi+\lambda_1 (\phi^\dagger\phi)^2+\lambda_2 (\chi^\dagger\chi)^2
+\lambda_3
(\chi^\dagger\chi)(\phi^\dagger\phi),%
\label{potential}%
\eea%
where $\lambda_3 > - 2 \sqrt{\lambda_1\lambda_2}$ and $\lambda_1,
\lambda_2 \geq 0$, so that the potential is bounded from below.
This is the stability condition of the potential. Furthermore, in
order to avoid vanishing vacuum expectation values (vevs):
$v=\langle \phi \rangle=0$ and $v'= \langle \chi \rangle=0$ from
being local minimum, one must assume that $\lambda_3^2 < 4
\lambda_1 \lambda_2$. As in the usual Higgs mechanism of
electroweak symmetry breaking in the SM, the $B-L$ spontaneous
symmetry breaking requires a negative squared masse, $m^2_{2} <
0$. In this case, the following non-zero
vev may be obtained%
\be %
v'^2 = \frac{-2 (m_1^2 +\lambda_1 v^2)}{\lambda_3}, %
\ee %
with %
\be%
v^2= \frac{4 \lambda_2 m_1^2 - 2 \lambda_3 m_2^2}{\lambda_3^2 -4
\lambda_1\lambda_2}.%
\ee %
From these equations, two comments are in order: $(i)$ For
non-vanishing $\lambda_3$, the vevs $v$ and $v'$ are related and
hence they are naturally of the same order, \ie, $v \simeq {\cal
O}(100)$ GeV and $v' \simeq {\cal O}(1)$ TeV. In fact, in this
scenario $v' \gg v$ will require a significant fine-tuning among
the input parameters: $m^2_{1,2}$ and $\lambda_{1,2,3}$. $(ii)$
The condition of the electroweak symmetry breaking, for $\lambda_3^2 -4 \lambda_1 \lambda_2 <0$, is given by%
\be %
m_1^2 < M^2_C = \frac{\lambda_3 m^2_2}{2 \lambda_2}.%
\ee %
For $m^2_{2} <0$ and $m_1^2 > M^2_C$, $U(1)_{B-L}$ is
spontaneously broken while the $SU(2)_L
\times U(1)_Y$ remains exact. At this stage, the following vevs are obtained %
\bea %
v' &=& \sqrt{\frac{-m_2^2}{2\lambda_2}},\\
v &=& 0 . \nonumber%
\eea %
The evolution from ${\cal O}(1)$ TeV scale down to ${\cal O}(100)$
GeV, may reduce the squared Higgs mass $m_1^2$ until eventually
the minimization condition is satisfied and the electroweak gauge
symmetry is broken. This scenario corresponds to two stages
symmetry breaking at two different scales. Note that if $\lambda_3
<0$, the radiative electroweak symmetry breaking can be achived
with positive squared Higgs mass. Therefore, throughout this work,
we will focus on the following region of mixing coupling
$\lambda_3$: $0 > \lambda_3 > -2 \sqrt{\lambda_1\lambda_2}$ and
$\lambda_{1,2} \sim O(1)$.

As usual, we expand the scalar field  $\chi$ around the $B-L$
minimum $v'$ and write %
\be%
\chi(x)=\frac{v^\prime+H^\prime(x)}{\sqrt{2}}. %
\ee %
In this case, one finds the following lagrangian for the $B-L$
Higgs ($H'$) mass and its interaction with
right handed neutrino and SM Higgs $\phi$: %
\be %
{\cal L}(\phi, H') = \frac{1}{2}m_{H^\prime}^2 H^{\prime 2} -
\frac {1}{2 \sqrt 2}\lambda_{\nu_R} H^\prime \bar{\nu}^c_R \nu_R +
\lambda_3 \left(\frac{1}{2} H^{\prime 2} \phi^2 +  v' H^\prime
\phi^2 \right). %
\label{Higgscoupling}%
\ee %

Finally, after the $B-L$ gauge symmetry breaking the gauge field
$C_{\mu}$ (will be called $Z^{\prime}$ in the rest of the paper) acquires the following mass: %
\be%
M_{Z^{\prime}}^2=4 g''^2 v'^2 .%
\label{mz}
\ee%
The Lagrangian terms that describe the interactions of the
$Z^\prime_{\mu}$ gauge boson are given by %
\be%
{\cal L}_{Z^{\prime}}=-\frac{1}{4}
Z^{\prime}_{\mu\nu}Z^{\prime^{\mu\nu}}+\frac{1}{2}
M_{Z^{\prime}}^2 Z^{\prime}_\mu Z^{\prime^\mu}+ 2g''^2
Z^{\prime}_\mu Z^{\prime^\mu}\chi^2+4 g''^2 v' Z^{\prime}_\mu
Z^{\prime^\mu} \chi - i g''J_\mu^{B-L} Z^{\prime^\mu}
\ee %
where $J_\mu^{B-L}= \bar{\psi}_L\gamma_\mu\psi_L +
\bar{e}_R\gamma_\mu e_R + \bar{\nu}_R \gamma^\mu \nu_R$. The high
energy experimental searches for an extra neutral gauge boson
impose lower bounds on the $Z^\prime$ mass. The LEP II provides
the most stringent constraint \cite{Carena:2004xs}. As $e^+ e^-$
collider, it was able to strongly constrain the extra-gauge boson
that coupled significantly with electrons. It is worth noting that
in this class of model, with the above particle assignments, any
mixing effects between $U(1)$ factors may arise only at the two
loop level, hence keeping them small enough. However, the
measurements of $e^+ e^- \to f \bar{f}$ above the $Z$-pole at LEP
II impose stringent constraints on the mass of $Z_{B-L}$ or on the
$B-L$ gauge coupling. Furthermore, the recent results by CDF II
\cite{Abulencia:2006iv} are consistent with the LEP II constraints
on $Z^{\prime}$ mass in case of $B-L$ extension of the SM.
Therefore, the typical lower
bound on $M_{Z^\prime}$ is now given by %
\be%
M_{Z^{\prime}} /g'' > 6 ~ \rm{TeV}. %
\ee%
Thus, one finds that $v' \gsim O(\rm{TeV})$.

\section{{\large{\bf Neutrino masses and mixing in $B-L$ extension of the
SM}}}
In this section we provide a detail analysis for the neutrino
masses and mixing in the gauge $B-L$ extension of the SM, where
the neutrino masses may be generated through a TeV scale seesaw
mechanism. After $U(1)_{B-L}$ symmetry breaking
\cite{Khalil:2006yi}, the Yukawa interaction term:
$\lambda_{\nu_R}\chi\bar{\nu}^c_{R}\nu_{R}$ leads, as usual, to
right handed neutrino mass: $ M_R = \frac{1}{2\sqrt
2}\lambda_{\nu_R} v' $. Also the electroweak symmetry breaking
results in the Dirac neutrino mass term :
$m_D=\frac{1}{\sqrt{2}}\lambda_\nu v$. Therefore, the mass matrix
of the left and right-handed neutrinos is
given by %
\be
\left(%
\begin{array}{cc}
  0 & m_D \\
  m_D & M_R \\
\end{array}%
\right). %
\ee%
Since $M_R$ is proportional to $v'$ and $m_D$ is proportional to
$v$ \ie, $M_R > m_D$, the diagonalization of this mass matrix
leads to the following masses for the light and heavy neutrinos respectively: %
\bea%
m_{\nu_L} &=& -m_D M_R^{-1} m_D^T ,\\
m_{\nu_H} &=& M_R . %
\eea %
Thus, $B-L$ gauge symmetry can provide a natural framework for the
seesaw mechanism. However, the scale of $B-L$ symmetry breaking
$v^{\prime}$ remains arbitrary. As in Ref.\cite{Khalil:2006yi},
$v^{\prime}$ is assumed to be of order TeV. Therefore, the value
of $M_R$ is also of that order.

In our analysis, we adopt a basis where the charged lepton mass
matrix and the Majorana mass matrix $M_R$ are both real and
diagonal.
Therefore, one can parameterize $M_R$ as follows%
\be%
M_R=M_{R_3}\left(%
\begin{array}{ccc}
  r_1 & 0 & 0 \\
  0 & r_2 & 0 \\
  0 & 0 & 1 \\
\end{array}%
\right), %
\label{MR}
\ee %
where %
\be %
M_{R_3}=\vert \lambda_{\nu_{R_3}} \vert
\frac{v'}{2\sqrt2}%
\ee %
and %
\be %
r_{1,2}=\frac{M_{R_{1,2}}}{M_{R_3}}=\left\vert
\frac{\lambda_{\nu_{R_{1,2}}}}{\lambda_{\nu_{R_3}}}\right\vert . \ee %
As can be seen from Eq.(\ref{MR}) that even if $v'$ is fixed to be
of order TeV, the absolute value of $M_R$ is still parameterized
by three known parameters. On the other hand, the Dirac mass
matrix (if it is real) is given in terms of $9$ parameters. Since
$U(1)_{B-L}$ can not impose any further constraint to reduce the
number of these parameters, the total number of free parameters
involved in the light neutrino mass matrix are $12$ parameters. As
is known, the solar and atmospheric neutrino oscillation
experiments have provided measurements for the neutrino
mass-squared differences and also for the neutrino mixing angles.
At the $3 \sigma $ level, the allowed ranges are
\cite{Altmann:2005ix} :
\begin{eqnarray}
\Delta m_{12}^2 &=& (7.9 \pm 0.4) \times 10^{-5} \rm{eV}^2 ,\nonumber\\
\vert \Delta m_{32}^2 \vert &=& (2.4+0.3)\times 10^{-3} \rm{eV}^2 ,\nonumber\\
\theta_{12} &=& 33.9^{\circ} \pm 1.6^{\circ}, \label{dm2}\\
\theta_{23} &=& 45^{\circ},\nonumber\\
\sin^2 \theta_{13} & \leq & 0.048 .\nonumber
\end{eqnarray}
Therefore, the number of the experimental inputs are at most six:
three neutrino masses (assuming possible ansatze like hierarchy or
degenerate) and three mixing angles (if we assume
$\theta_{13}=0$).

One of the interesting parametrization for the Dirac
neutrino mass matrix is given by%
\be %
m_D=U_{MNS}\sqrt{m_\nu^{diag}}R\sqrt{M_R}    , %
\label{mD}
\ee %
where $m_\nu ^{diag}$ is the physical neutrino mass matrix and
$U_{MNS}$ is the lepton mixing matrix. The matrix R is an
arbitrary orthogonal matrix which can be parameterized, in case of
real $m_D$, in terms of three angles. In Eq.(\ref{mD}), the six
unknown parameters are now given in terms of three masses in $M_R$
and the three angles in $R$. In order to fix these angles, one
needs a flavor symmetry beyond the gauge symmetry which is
typically flavor blind. Several types of flavor symmetries have
been discussed in the literatures \cite{Mohapatra:2006gs}. Here we
follow different approach. We attempt to extend the observed
relations between the masses of up quarks and charged leptons to
the down quark and neutrino masses.

From the measured values of the up quark and charged lepton masses
at the electroweak scale, one can notice the following relations %
\be %
\frac {m_u}{m_c}\sim \frac {m_e}{m_\mu}\sim O(10^{-3}),%
\ee %
and %
\be %
\frac {m_c}{m_t}\sim \frac {{m_\mu}^2}{{m_\tau}^2}\sim
O(10^{-3}) %
\ee %
In the event of a flavor discrete-symmetry that may explain these
ratios, the down quark and neutrino sectors may also be subjected
to this symmetry. Hence, a similar relation may be obtained among
their masses. If the scale of this discrete symmetry ($v_F$) is
below seesaw ($B-L$ symmetry breaking) scale, then the above mass
ratio would be extended to the down quark and light neutrino
masses.
In this case, one would expect that %
\be %
\frac{m_d}{m_s}\sim \frac{m_{{\nu}_1}}{m_{\nu_2}} \sim  O(10^{-2})%
\label{down1}
\ee%
\be %
\frac{m_s}{m_b}\sim \frac{m^2_{\nu_2}}{m^2_{\nu_3}}\sim
O(10^{-2}).%
\label{down2}
\ee %
However, if the scale of the flavor discrete-symmetry is above the
seesaw mechanism scale, then the mass ration is anticipated to be
between down quark and Dirac neutrino masses, \ie,
\be %
\frac{m_d}{m_s}\sim \frac{m_{D_1}}{m_{D_2}} \sim  O(10^{-2})%
\label{down3}
\ee%
\be %
\frac{m_s}{m_b}\sim \frac{m^2_{D_2}}{m^2_{D_3}}\sim
O(10^{-2})%
\label{down4}
\ee %

Let us start by considering the first scenario where $v_F <
v^\prime$. The experimental results for the light neutrino
masses in Eq.(\ref{dm2}) leads to %
\bea%
m_{\nu_2}&=& \sqrt{7.9\times 10^{-5} - m_{\nu_1}^2},\\
m_{\nu_3}&=& \sqrt{\vert 2.4\times 10^{-3} - 7.9\times 10^{-5} +
m_{\nu_1}^2 \vert},%
\eea%
with arbitrary $m_{\nu_1}$. Thus, for $m_{\nu_1}^2 \ll 7.9 \times
10^{-5}$, the ansatz of hierarchal light neutrino masses is
obtained. It is interesting to note that if $m_{\nu_1} \sim
10^{-4}$, the hierarchal ansatz is consistent with the mass
relations given in Eqs.(\ref{down1},\ref{down2}) and the light
neutrino mass matrix takes the form %
\be%
m_\nu \simeq 0.05~ eV \left(%
\begin{array}{ccc}
  10^{-3} & 0 & 0 \\
  0 & 0.16 & 0 \\
  0 & 0 & 1 \\
\end{array}%
\right)~.%
\ee %
The Dirac neutrino mass matrix is now given by (for $r_1 \sim r_2
\sim 0.1$, $M_{R_3}= 5$ TeV, and order one angles/phases of
$R$-matrix):
\be %
m_D \simeq 10^{-3}~ \left(%
\begin{array}{ccc}
  0.16 +0.23~ i & -0.25 + 0.16~ i  & -0.19 - 0.26~ i \\
  -0.22 - 0.34~ i & 0.37 -0.24~ i  & 0.30 + 0.38~ i \\
  -0.14 + 0.47~ i & -0.53 - 0.15~ i  & 0.16 - 0.68~ i \\
\end{array}%
\right).%
\ee%
Note that the complex phases in $m_D$ are induced by the phases of
$R$ matrix since the mixing matrix $U_{MNS}$ is real
($\theta_{13}=0$ is assumed). These phases are crucial for
generating lepton asymmetry as will be discussed in the next
section. Also, as can be seen from the above example, $m_D  \lsim
{\cal O}(10^{-4})~ \rm{GeV}$, \ie, the Dirac neutrino Yukawa
coupling $\lambda_{\nu}$ is of order $10^{-6}$, which is just one
order of magnitude smaller than the electron Yukawa coupling.

Now we turn to the case of $v_F > v'$. From Eqs. (\ref{down3},\ref{down4}), one gets %
\be%
m_D^{diag}\simeq
m_{D_3}\left(%
\begin{array}{ccc}
  10^{-3} & 0 & 0 \\
  0 & 10^{-1} & 0 \\
  0 & 0 & 1 \\
\end{array}%
\right). \label{mDdiag}%
\ee%
If we assume hierarchal neutrino masses $m_{\nu_1} \ll m_{\nu_2}
\ll m_{\nu_3}$, the light neutrino mass matrix can be written as %
\be%
m_\nu \simeq 0.05~ eV \left(%
\begin{array}{ccc}
  m_{\nu_1} & 0 & 0 \\
  0 & 0.16 & 0 \\
  0 & 0 & 1 \\
\end{array}%
\right)%
\ee %
By using the determinant of $m_D$ from Eqs.(\ref{mD}) and
(\ref{mDdiag}), one can express $m_{D_3}$ in terms of $m_{\nu_1}$,
$r_1$,
$r_2$ and $M_{R_3}$ as follows: %
\be %
\left(\frac{m_{D_3}}{\rm{GeV}}\right) \simeq 10^{-4}~
\left(\frac{M_{R_3}}{\rm{GeV}}\right)^{1/2} ~ \left[r_1~ r_2 \left(\frac{m_{\nu_1}}{\rm{eV}}\right)\right]^{1/6} ~ .%
\ee %
Here, we have used the fact that the determinant of the orthogonal
matrix $R$ is one. Using this relation, one can determine, in
terms of $M_{R_3}$, $r_1$ and $r_2$, the three angles
($\theta_{12},\theta_{23},\theta_{13}$) that parameterize the
matrix $R$ and lead to eigenvalues for the Dirac mass matrix $m_D$
consistent with our inputs in Eq.(\ref{mDdiag}).

In case of $r_1 \lsim r_2 \lsim 1$ (hierarchy heavy neutrino
masses), one finds that there is a possible solution for the
angles $\theta_{ij}$ only for $m_{\nu_1}< 10^{-7}$ GeV. In
addition the angle $\theta_{13}$ can be fixed at
$\theta_{13} \simeq \pi/2$, hence the matrix $R$ is given by%
\be %
R=\left(%
\begin{array}{ccc}
 0 & 0 & 1 \\
 -\sin\alpha & \cos\alpha& 0 \\
  -\cos\alpha & -\sin\alpha & 0 \\
\end{array}%
\right),%
\ee %
where $\alpha= \theta_{12}+\theta_{23}$. For instance, with
$M_{R_3}= 5$ TeV, $r_1\simeq 0.1$ and $r_2\simeq 0.4$,
one gets $\alpha \simeq 0.65$. Thus, the following $R$ matrix is obtained%
\be %
R= \left(%
\begin{array}{ccc}
  0 & 0 & 1 \\
  -0.6 & 0.8 & 0 \\
  -0.8& -0.6 & 0 \\
\end{array}%
\right).%
\ee%
While, for $r_1\simeq r_2 \simeq 1$ (degenerate heavy neutrino
masses), the matrix $R$ is given by
\be %
R= \left(%
\begin{array}{ccc}
  0 & 0 & 1 \\
  -0.73 & 0.67 & 0 \\
 -0.67 & -0.73 & 0 \\
\end{array}%
\right).%
\ee%
Finally, we can also have a complex R, which induce a new source
of CP violation phase in the Dirac Yukawa matrix $Y_D$. In this
case, the angle $\alpha$ would be written as $\alpha = \rho + i
\sigma$. For the above example of $r_1=0.1$ and $r_2=0.4$, the
corresponding complex $R$-matrix is given by %
\be %
R= \left(%
\begin{array}{ccc}
 0 & 0 & 1 \\
  -0.6~ e^{i a} & 0.8~ e^{i a}& 0 \\
  -0.8 ~ e^{i a} & -0.6~ e^{i a} & 0 \\
\end{array}%
\right).%
\ee%
It is important to mention that the complex phases in $R$ matrix
are not related to any of the low energy phases, however, it plays
a crucial role in leptogenesis .

Before closing this section, we comment on the scenario of
degenerate light neutrino masses ($m_{\nu_1}\simeq m_{\nu_2}\simeq
m_{\nu_3}\simeq \tilde{m}$). From the astrophysical constraint:
$\sum_i m_{\nu_i} < 1
~\rm{eV}$, one finds that $\tilde{m} < 0.3$ eV. In this case, one may write %
\bea%
m_{\nu_2}&=& \sqrt{\tilde{m}^2 + 7.9\times 10^{-5}},\\
m_{\nu_3}&=& \sqrt{\tilde{m}^2+ 2.4\times 10^{-3} + 7.9\times
10^{-5}}.%
\eea%
Therefore, the light neutrino mass matrix takes the form %
\be %
m_{\nu} = \tilde{m}~ \left(%
\begin{array}{ccc}
  1 &   &  \\
  & \sqrt{1+ \frac{0.000079}{\tilde{m}^2}} & \\
  &   & \sqrt{1+ \frac{0.002479}{\tilde{m}^2}} \\
\end{array}%
\right) \lsim  0.3~\rm{eV}~ \left(%
\begin{array}{ccc}
  1 &   &  \\
  & 1.00044 & \\
  &   & 1.01368 \\
\end{array}%
\right).%
\ee%
In this case, it is clear that the suggested mass relations
between down type quark and neutrino masses should be implemented
on the Dirac neutrino masses. However, we found that there is no
any solution for the angles $\theta_{ij}$ that can account for
$m_D$. Therefore, in our framework, the ansatz of degenerate
neutrino masses is disfavored .

\section{{\large{\bf TeV scale Leptogenesis in $B-L$ extension of the
SM}}}
The recent observations indicate that the asymmetry between number
density of baryon ($n_B$) and of anti-baryon ($n_{\bar{B}}$) of
the universe is given by \cite{Spergel:2003cb}
\be%
Y_B= \frac{n_B -n_{\bar{B}}}{s}= \frac{n_B}{s} = (6.3 \pm 0.3) \times 10^{-10},%
\label{YBresult}
\ee%
where $s=2\pi^2 g_* T^3/45$ is the entropy density and $g_*$ is
the effective number of relativistic degrees of freedom.

\begin{figure}[h,t]
\begin{center}
\epsfig{file=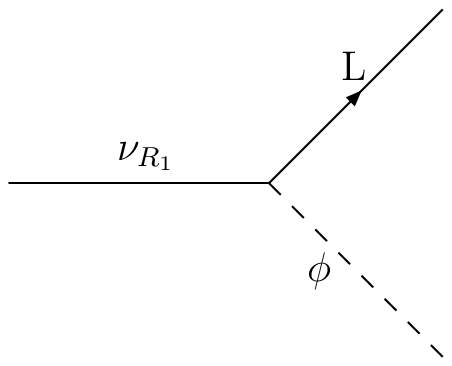,
width=4.5cm,height=3.25cm,angle=0}~~~~~\epsfig{file=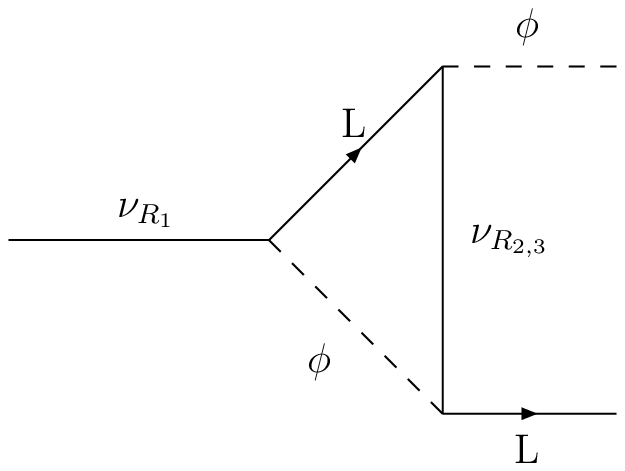,
width=4.5cm,height=3.25cm,angle=0}~~~~~\epsfig{file=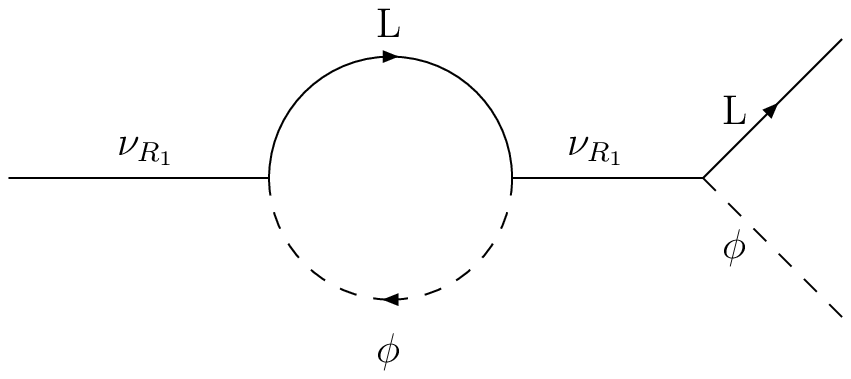,
width=5.5cm,height=2.5cm,angle=0}
\end{center}
\caption{Feynman diagrams in SM with right-handed neutrinos that
contribute to the decay $\nu_{R_1} \to \phi~ l_{\alpha}$.}
\label{fig:smdiagram}
\end{figure}

\begin{figure}[h,t]
\begin{center}
\epsfig{file=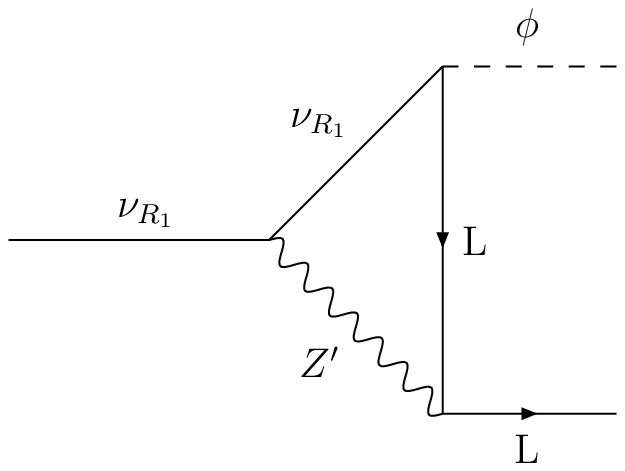,
width=5.5cm,height=3.5cm,angle=0}~~~~~~\epsfig{file=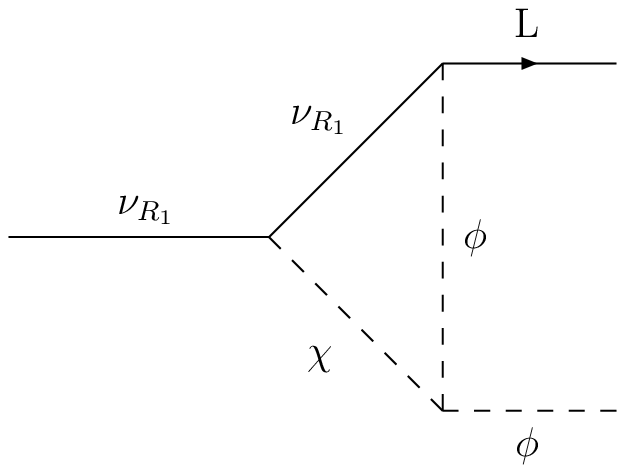,
width=5.5cm,height=3.5cm,angle=0}
\end{center}
\caption{The new contributions to the decay $\nu_{R_1} \to \phi~
l_{\alpha}$ in $B-L$ extension of the SM.} \label{fig:B-Ldiagram}
\end{figure}

As mentioned in the introduction, the possibility of originating
this asymmetry through the CP violating decay of the heavy
right-handed neutrino is an interesting mechanism known as
Leptogenesis \cite{Chen:2007fv}. Within the framework of $B-L$
extension of the SM, the lepton asymmetry $\varepsilon_i$ is
generated by the CP violating decays of $\nu_{R_i}$ into the Higgs
doublet and the charged lepton doublet $l_{\alpha}$, \ie,
$\nu_{R_i} \to \phi + l_{\alpha}$ where $\alpha=(e, \mu, \tau)$.
The lepton asymmetry is usually dominated by the $\nu_{R_1}$
decay:
\be %
\varepsilon_1 = \frac{\sum_{\alpha} \left(\Big\vert A(\nu_{R_1}
\to \phi~ l_{\alpha})\Big\vert^2 - \Big\vert A(\nu_{R_1} \to
\bar{\phi}~ \bar{l}_{\alpha})\Big\vert^2 \right)}{\sum_{\alpha}
\left(\Big \vert A(\nu_{R_1} \to \phi~ l_{\alpha})\Big\vert^2 +
\Big\vert A(\nu_{R_1} \to \bar{\phi}~
\bar{l}_{\alpha})\Big\vert^2 \right)},%
\ee %
where $A(\nu_{R_1} \to \phi~ l_{\alpha})$ is the total (tree plus
loop) decay amplitude. Similar to the SM extended by three right
handed neutrinos, the decay of $\nu_{R_1}$ into $\phi$ and
$l_{\alpha}$ may occur through the tree level diagram, one loop
vertex correction, and one loop self-energy, as shown in Fig.
\ref{fig:smdiagram}.
However, in $B-L$ extension of the SM the decay of the
right-handed neutrino into Higgs and the lepton doublets can be
also generated through diagrams mediated by extra Higgs and extra
gauge boson exchange, as displayed in Fig. \ref{fig:B-Ldiagram}.

In order to analyze the CP asymmetry $\varepsilon_1$, one may
parameterize the decay amplitude as CP violating part times a CP
conserving part (comes mainly from the loop function). In this
respect, one may write tree level $A_0(\nu_{R_1} \to \phi~
l_{\alpha})$ and one loop level
$A_1(\nu_{R_1} \to \phi~ l_{\alpha})$ as follows: %
\bea %
A_0(\nu_{R_1} \to \phi~ l_{\alpha}) &=& A_{tree}, ~~~~~~~~~~~~~~~
\bar{A}_0(\nu_{R_1} \to \phi~ l_{\alpha})= A^*_{tree},\\
A_1(\nu_{R_1} \to \phi~ l_{\alpha}) &=& A_{loop} \times F,
~~~~~~~~~
\bar{A}_0(\nu_{R_1} \to \phi~ l_{\alpha})= A^*_{loop} \times F. %
\eea %
The CP asymmetry arises through the interference between tree and
loop contributions, hence $\varepsilon_1$ can be written as%
\be%
\varepsilon_1 = \frac{{\rm Im}\left[A_{tree} A^*_{loop}\right]
~{\rm Im}\left[F\right]}{\vert A_{tree}\vert^2 + \vert A_{loop}~F
\vert^2 + 2 {\rm Re}[F]~ {\rm Re}\left[A_{tree}~A^*_{loop}\right]}, %
\ee %
Since $\vert A_{tree} \vert \gg \vert A_{loop}~F \vert $, the CP
asymmetry
$\varepsilon_1$ is usually approximated as %
\be%
\varepsilon_1 \simeq \frac{1}{\vert A_{tree}\vert^2}~ {\rm
Im}\left[A_{tree} A^*_{loop}\right] ~{\rm Im}\left[F\right]. %
\ee %

The tree level contribution to the $\nu_{R_1}$ decay amplitude is
given by
\be %
A_0(\nu_{R_1} \to \phi~ l_{\alpha}) = -i (\lambda_{\nu})_{\alpha
1}
\Big(\bar{u}(p)~ P_R~ u^c(q)\Big), %
\ee %
where $\bar{u}$ is the Dirac spinor of outgoing particle
$l_{\alpha}$ with momentum $p$ and $u^c(q)$ is the spinor of the
ingoing particle $\nu_{R_1}$ with momentum $q$. As can be seen
from the above expression, the tree level contribution is
proportional to Dirac neutrino Yukawa couplings
$(\lambda_{\nu})_{\alpha1}$ which is of order $10^{-6}$, hence it
is quite small. The contribution to the decay
amplitude from the vertex correction is given by %
\be %
A_{V}(\nu_{R_1} \to \phi~ l_{\alpha}) = \frac{i}{16 \pi^2} \sum_i
(\lambda_{\nu})_{\alpha i} (\lambda^{\dagger}_{\nu}
\lambda_{\nu})_{1i}
\Big(\bar{u}(p)~ P_R~ u^c(q)\Big)~ F_V\left(\frac{M_{R_i}^2}{M_{R_1}^2}\right). %
\ee %
The loop function  $F_V(x)$ is given by %
\bea %
{\rm Re}~F_V(x) &=& -\sqrt{x}\left[1+\ln(x)[1-(1+x)
\ln\left(\frac{1+x}{x}\right)\right], \\ %
{\rm Im}~F_V(x) &=& \sqrt{x}\left[1-(1+x)
\ln\left(\frac{1+x}{x}\right)\right]. %
\eea %
The self-energy diagram leads to the following
contribution to the decay amplitude of $\nu_{R_1} \to \phi~  l_{\alpha}$ %
\be %
A_{S}(\nu_{R_1} \to \phi~ l_{\alpha}) = \frac{i}{16 \pi^2} \sum_i
(\lambda_{\nu})_{\alpha i} (\lambda^{\dagger}_{\nu}
\lambda_{\nu})_{1i}
\Big(\bar{u}(p)~ P_R~ u^c(q)\Big)~ F_S\left(\frac{M_i^2}{M_{R_1}^2}\right), %
\ee %
where the corresponding loop function $F_S(x)$ is given by
\cite{buchmuler}
\be %
F_S\left(\frac{M_{R_i}^2}{M_{R_1}^2}\right)= \large\vert\omega_{1i}(M_{R_1}^2)\large\vert^2 (M_{R_1}^2-M_{R_i}^2)\frac{M_{R_i}}{M_{R_1}}. %
\label{self energy}\ee %
with
\be%
\omega_{1i}(M_{R_1}^2)^{-1}=\frac{(M_{R_1}^2-M_{R_i}^2)}{M_{R_1}}-2a(M_{R_1}^2)\left(M_{R_1}
(\lambda_\nu^\dagger\lambda_\nu)_{ii}-M_{R_i}(\lambda_\nu^\dagger\lambda_\nu)_{11}\right)
\label{omega}\ee and
$$a(q^2)=\frac{1}{16 \pi^2}\left(\ln \frac{q^2}{\mu^2}-2-i\pi \Theta(q^2)\right).$$
It is clear that the loop function $F_S$ approaches zero in case
of degenerate right-handed neutrino masses and hence the amplitude
$A_S$ vanishes identically. In our model with TeV scale seesaw
mechanism, the Dirac neutrino Yukawa couplings
$(\lambda_\nu)_{ij}$ are of order $10^{-6}$. Therefore, the second
term in $w_{1i}^{-1}$ is much smaller than the first term, unless
the right-handed neutrino masses are completely degenerate. In
this
respect, the loop function $F_S$ in Eq.(\ref{self energy}) is reduced to%
\be%
F_S\left(\frac{M_{R_i}^2}{M_{R_1}^2}\right)=\frac{M_{R_1}
M_{R_i}}{M_{R_1}^2-M_{R_i}^2}%
\ee

Now we turn to the new $B-L$ contributions to the right-handed
neutrino decay due to the exchanges of extra $Z'$ and $H'$ as
shown in Fig. \ref{fig:B-Ldiagram}. It is expected that these
diagrams do not include strong CP phases since the running
particles in the loops are heavier than initial particle
$\nu_{R_1}$ and hence the these diagrams can not be cut in two
parts. As a result, these diagrams will not directly contribute to
the lepton asymmetry. However, These contributions may be of order
the tree level and hence they affect the interference with vertex
and self-energy corrections that remain the only sources of strong
CP phases.

Our result for the extra Higgs contribution to the decay amplitude
of $\nu_{R_1} \to \phi~  l_{\alpha}$ leads to %
\be %
A_{\chi}(\nu_{R_1} \to \phi~ l_{\alpha}) = \frac{1}{16 \pi^2
M_{R_1}} (\lambda_{\nu_R} )_{11} (\lambda_{\nu})_{1\alpha}~
g_{\phi^2 \chi}~ \left(\bar u(p)~ P_R~ u^c(q)\right)~ F_{\chi}\left(\frac{M_{\chi}^2}{M_{R_1}^2}\right). %
\ee %
From Eq.(\ref{Higgscoupling}), the coupling $g_{\phi^2 \chi}$ is given by :%
\be %
g_{\phi^2 \chi} = \sqrt{2} v^\prime \lambda_3. %
\label{gH}
\ee %
As can be seen from Eq.(\ref{potential}), the mixing parameter
$\lambda_3$ is real thus the coupling $g_{\phi^2\chi}$ is real.
Moreover, as explained in the previous section, the Yukawa
coupling $(\lambda_{\nu_R})_{11}$ is real too. In
general it has the form: %
\be %
(\lambda_{\nu_R})_{11} = \frac{2 \sqrt{2} M_{R_1}}{v^\prime}. %
\label{Mphase}
\ee %

Finally, $F_{\chi}(x)$ is the associate loop function which is  given by %
\bea %
F_{\chi}(x) &=&
1-\frac{\pi^2}{6}-\left(\frac{x}{2}+\ln(2)-1\right)\ln(x)-
\frac{1}{2}\sqrt{x(x-4)}\ln\left(\frac{x-\sqrt{x(x-4)}}{x+\sqrt{x(x-4)}}\right) \nonumber\\
&-&\ln\left(x-\sqrt{x(x-4)}\right)\ln\left(\frac{x-\sqrt{x(x-4)}-2}{x-\sqrt{x(x-4)}}\right)\nonumber\\
&-&\ln\left(x+\sqrt{x(x-4)}\right)\ln\left(\frac{x+\sqrt{x(x-4)}-2}{x+\sqrt{x(x-4)}}\right)\nonumber\\
&+&Li_2\left(\frac{-x+\sqrt{x(x-4)}+2}{-x+\sqrt{x(x-4)}}\right)+Li_2\left(\frac{x+\sqrt{x(x-4)}-2}{x+\sqrt{x(x-4)}}\right)
\eea %

Our computation for the the extra gauge boson $Z^\prime$
contribution to the decay amplitude of $\nu_{R_1} \to \phi~  l_{\alpha}$ leads to %
\be %
A_{Z^{\prime}}(\nu_{R_1} \to \phi~ l_{\alpha}) = \frac{1}{4 \pi^2}
g^{''^2} (\lambda_{\nu})_{\alpha 1} \Big(\bar u(p) P_R u^c(q)\Big)~ F_{Z^{\prime}}\left(\frac{M_{z'}^2}{M_{R_1}^2}\right), %
\ee %
where $F_{Z^\prime}(x)$, at the $M_{R_1}$ scale, is given by%
 \bea%
F_{Z^{\prime}}(x)&=&
-\frac{3}{2}+\frac{\pi^2}{12}-\sqrt{-x(x-4)}(x+1)\left[\tan^{-1}\left(\frac{x-2}{\sqrt{-x(x-4)}}\right)
+\tan^{-1}\left(\frac{\sqrt{-x}}{\sqrt{(x-4)}}\right)\right]\nonumber\\
&+& \left(1+\frac{1}{4}x+x\ln(4)-\frac{1}{2}(1-4x)\ln
2\right)\ln(x)-2\pi^2 x\nonumber\\
&+&-
(\frac{1}{4}+x)\sqrt{x(x-4)}\ln\left(\frac{-\sqrt{x(x-4)}+x}{\sqrt{x(x-4)}+x}\right)\nonumber\\
&+&
2x\ln\left[\frac{x+\sqrt{x(x-4)}}{x}\right]\ln\left(\frac{-2+x-\sqrt{x(x-4)}}{x-\sqrt{x(x-4)}}\right)\nonumber\\
&+& 2x\ln\left[\frac{x-\sqrt{x(x-4)}}{x}\right]\ln\left(\frac{-2+x+\sqrt{x(x-4)}}{x+\sqrt{x(x-4)}}\right) \nonumber\\
&-&\frac{1}{2}(1-4x)\ln\left[x+\sqrt{x(x-4)}\right]\ln\left(\frac{-2+x+\sqrt{x(x-4)}}{x+\sqrt{x(x-4)}}\right)\nonumber\\
&-&\frac{1}{2}(1-4x)\ln\left[x-\sqrt{x(x-4)}\right]\ln\left(\frac{-2+x-\sqrt{x(x-4)}}{x-\sqrt{x(x-4)}}\right)\nonumber\\
&+&\frac{1}{2}Li_2\left(\frac{1}{2}(1-\frac{\sqrt{(x-4)}}{\sqrt{x}})\right)+\frac{1}{2}Li_2\left(\frac{1}{2}(1+\frac{\sqrt{(x-4)}}{\sqrt{x}})\right)
\eea%

From the above expressions, it can be easily noted that the CP
violating effect in the amplitudes $A_V(\nu_{R_1} \to \phi~
l_{\alpha})$ and $A_S(\nu_{R_1} \to \phi~ l_{\alpha})$ arises from
the same source: $(\lambda_{\nu})_{\alpha i}
(\lambda^+_{\nu}\lambda_{\nu})_{1i}$. While in $A_{\chi}(\nu_{R_1}
\to \phi~ l_{\alpha})$ and $A_{Z^\prime}(\nu_{R_1} \to \phi~
l_{\alpha})$ they are proportional to $g_{\phi^2\chi}
(\lambda_{\nu_R})_{11} (\lambda_{\nu})_{1\alpha}$ and $g^{''^2}
(\lambda_{\nu})_{\alpha 1}$, respectively. Therefore, if
$\lambda_{\nu} \sim \mathcal{O}(10^{-6})$, as found in previous
section, the $\chi$ and $Z^\prime$ may give significant
contributions. However, due to the absence of strong CP violation
in these processes, they have no interference with tree level
diagram. Nevertheless, they may have significant effect through
the interference with the SM one loop amplitudes $A_V$ and $A_S$.
In this case, the total asymmetry is given by:
\bea%
\varepsilon_1 & \simeq & \frac{1}{8\pi(\lambda_{\nu}^\dagger
\lambda_{\nu})_{11}}\sum_{i=2,3}\Big[ \frac{ {\rm
Im}\{(\lambda_{\nu}^\dagger
\lambda_{\nu})^2_{1i}\}}{\Big(1-\frac{1}{4\pi^2}g''^2
F_{Z^{\prime}}\left(\frac{M_{z'}^2}{M_{R_1}^2}\right)-\frac{1}{4\pi^2}\lambda_3
\ F_{\chi}\left(\frac{M_{\chi}^2}{M_{R_1}^2}\right)\Big)}
\Big]\nonumber\\& &\left[{\rm
Im}~F_V\left(\frac{M_{R_i}^2}{M_{R_1}^2}\right) +
{\rm Im}~F_S\left(\frac{M_{R_i}^2}{M_{R_1}^2}\right)\right].%
\eea

If the new contributions due to $\chi$ and $Z^\prime$ exchanges
have been neglected, one gets the usual CP asymmetry
$\varepsilon_1$ of the SM extended by right handed neutrinos,
which is given by
\be%
\varepsilon^{\rm{SM}}_1 \simeq \frac{1}{8\pi}
\frac{1}{(\lambda_{\nu}^\dagger\lambda_{\nu} )_{11}} \sum_{i=2,3}
{\rm Im}\{(\lambda_{\nu}^\dagger\lambda_{\nu} )^2_{1i}\}
\left[{\rm Im}~F_V\left(\frac{M_{R_i}^2}{M_{R_1}^2}\right) +
{\rm Im}~F_S\left(\frac{M_{R_i}^2}{M_{R_1}^2}\right)\right].%
\ee %
From the equations, few comments are in order: $(i)$ The lepton
asymmetry obtained in SM with right handed neutrinos is sensitive
to the CP phase of $(\lambda_{\nu} \lambda_{\nu}^\dagger)$.
Therefore, the necessary condition for the mechanism of
leptogenesis to work is %
\be %
{\rm Im}\left( \lambda_{\nu}^\dagger\lambda_{\nu}\right)_{1i}\neq
0 \Rightarrow {\rm Im}\left(\sqrt{M_R}~ R ~ m_{\nu}^{\rm diag} R^+
\sqrt{M_R}\right) \neq 0, ~ i=2,3. %
\ee %
$(ii)$ Due to the unitarity of the $U_{\rm MNS}$, leptogenesis
does not depend on the phases (if any) appearing in the leptonic
mixing matrix. $(iii)$ If the matrices $R$ and $M_R$ are real,
then $\varepsilon_1=0$ and hence the leptogenesis vanishes
identically. $(vi)$ In the limit of quasi-degenerate right-handed
neutrinos \ie, $x =(M_{R_2}/M_{R_1})^2\sim 1$, an enhancement for
$\varepsilon_1$, due to the nearly vanishing of the denominator of
$F_S(x)$, is obtained \cite{resonant}. $(v)$ A possible
enhancement for $\varepsilon_1$ can be achieved if
$\frac{g^{''}}{4\pi} F_{Z'} \simeq 1$. Note that since $\lambda_3
< 1$, the $H'$ contribution is typically smaller than the tree
level one.

\subsection{{\large{\bf Thermal leptogenesis}}}

This is the simplest scenario for leptogenesis where the lightest
right-handed neutrino, $\nu_{R_1}$, is assumed to be in
equilibrium while the heavier ones are decaying
\cite{Fukugita:1986hr} . In this respect, the leptogenesis can be
realized by the out of equilibrium of $\nu_{R_1}$ at temperature
below its mass scale. To avoid washing out the asymmetry
$\varepsilon_1$ by inverse decay and scattering processes, the
total width of $\nu_{R_1}$ decay should be smaller than the
expansion rate of the universe at temperature $T=M_{R_1}$. This is
known as
out-of-equilibrium condition, which implies that \cite{Fukugita:1986hr}%
\be %
Y_L = \frac{n_L - n_{\bar{L}}}{s} = \eta~
\frac{\varepsilon_1}{g_*},%
\ee %
where $\eta$ is the efficiency factor which parameterizes the
amount of washing out which depends on the size of
$r=\Gamma_1/H(M_{R_1}) \simeq m_{\nu_1}/m^*$ where $m^*= 256
\sqrt{g_*} v^2/3 M_P$. If $r\ll 1$ \ie, $\nu_{R_1}$ decays
strongly out-of-equilibrium, then $\eta \sim 1$. For $ r\gg 1$,
the lepton asymmetry is suppressed by $\eta \simeq 1/r$. Finally.
the electroweak sphaleron effects convert the lepton asymmetry
$Y_L$ to baryon asymmetry $Y_B$ through a
conversion factor $c$ \cite{Fukugita:1986hr}:%
\be %
Y_B = \frac{c}{c-1} Y_L \simeq -1.4 \times 10^{-3} \eta~
\varepsilon_1. %
\ee %
\begin{figure}[h,t]
\begin{center}
\epsfig{file=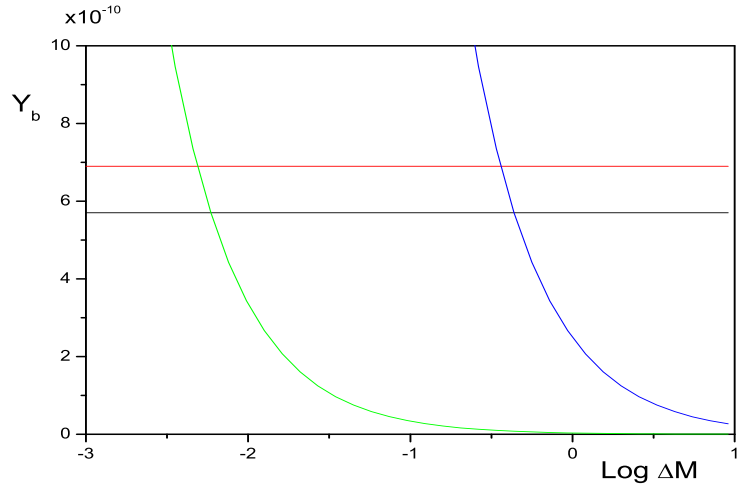, width=11cm,height=7.5cm,angle=0}
\end{center}
\caption{Baryon asymmetry in $B-L$ extension of the SM versus the
mass difference of the first two right-handed neutrinos.
Horizontal lines indicate the allowed $2 \sigma$ region of $Y_B$.}
\label{deltam2}
\end{figure}
From this expression, it is clear that $Y_B$ can be of order the
observed results reported in Eq.(\ref{YBresult}) only if
$\varepsilon_1$ is of order ${\cal O}(10^{-6})$, assuming $\eta
\sim 1$. However, as mentioned, unless the masses of the first two
right-handed neutrinos are quite degenerate and/or the $Z'$
contribution is of order the tree level one, the lepton asymmetry
is a few order of magnitude below this value.

The baryon asymmetry as function of the mass difference $\Delta M
= M_{R_2}- M_{R_1}$ is shown in Fig.\ref{deltam2}. As can be seen
from this figure, In case of negligible $Z'$ contribution, one
needs $\Delta M\simeq {\mathcal O}(10^{-3})$ to have $Y_B$ within
the $2 \sigma$ range of the experimental measurements. While, for
large $Z'$ contribution, this degeneracy constrain is relaxed and
mass difference of order $10\%$ can account for the observed
baryon asymmetry.

\subsection{{\large{\bf Non-thermal leptogenesis}}}
Now we consider the possibility of having non-thermal leptogenesis
\cite{Asaka:1999yd}.  As mentioned in the introduction, in the
non-thermal leptogenesis scenario the right-handed neutrino is a
decay product of a heavier particle. It is interesting to note
that in our model of $B-L$, the extra Higgs $\chi$ and extra gauge
boson $Z^\prime$ have direct couplings with the right handed
neutrino $\nu_{R_1}$. In case, of extra-Higgs, this coupling is
given (see Eq.\ref{Higgscoupling}) by
$-\lambda_{\nu_{R_1}}/2\sqrt{2}$. Therefore, the decay $\chi \to
\nu_{R_1} \nu_{R_1}$ is kinematically allowed if $m_{\chi} > 2
M_{R_1}$. In this case, one finds the following decay rate
$\Gamma_\chi$:%
\be %
\Gamma_\chi = \Gamma(\chi \to \nu_{R_1} \nu_{R_1}) = \frac{1}{4
\pi} \vert \lambda_{\chi \nu_{R_1}} \vert^2 M_{\chi}= \frac{1}{4
\pi} \frac{M_{R_1}^2}{v'^2} M_{\chi}. %
\ee%
The reheating temperature after this decay is given by %
\be %
T_R = \left(\frac{45}{4 \pi^3 g_*}\right)^{1/4} \left( \Gamma_\chi
M_P \right)^{1/2}, %
\ee %
where $M_P$ is Planck mass. In this framework, the lepton
asymmetry is given by \cite{Asaka:1999yd}
\be %
Y_L = \frac{3}{2} BR(\chi \to \nu_{R_1} \nu_{R_1})
\frac{T_R}{M_{\chi}} \varepsilon_1. %
\ee %
In order to avoid the inverse decay of $\chi \to \nu_{R_1}
\nu_{R_1}$, the decay rate should be less than the universe
expansion rate, \ie, $\Gamma(\chi \to \nu_{R_1} \nu_{R_1}) \lsim
H(\frac{M_{\chi}}{M_P})$, where $H(M_{\chi}) \sim 1.7 \sqrt{g_*}
M_{\chi}^2 /M_P$ with $g_{*}=100$ in the SM. This out of
equilibrium condition imposes stringent constrain on the coupling
$\lambda_{\chi\nu_{R_1}}$ and/or on the extra-Higgs mass. For
instance, if $M_{\chi} \simeq {\mathcal O}(100)$ TeV then
$\lambda_{\chi \nu_{R_1}}$ should be $\lsim 10^{-5}$ and the
associated reheating temperature is of order $\lsim 10^{7}$ GeV.
Therefore, the factor $T_R/M_\chi$ is of order $10^{2}$. In this
case, the the lepton asymmetry is enhanced by two order of
magnitude at most. It is clear that this enhancement alone is not
enough to account for the measured baryon asymmetry in the
universe. However, it helps in relaxing the condition of nearly
degenerate right-handed neutrino masses.

%
\section{{\large{\bf Conclusions}}}
In this paper we have systematically analyzed the phenomenological
implications for TeV scale $B-L$ extension of the SM. We have
investigated the possible scenarios of symmetry breaking and the
consequence on low energy experiments. We have studied the
neutrino masses and mixing in this type of models. We have shown
that the low scale seesaw mechanism is naturally implemented.
However, to fix the free parameters of the neutrino sector and
determine the Dirac neutrino mass matrix, a kind of flavor
symmetry is required. We assumed a phenomenological mass relation
between quarks and leptons. In this respect, we found that a
hierarchal ansatz for light neutrino masses is favored.

We have also analyzed the leptogenesis in this class of models. We
computed the new contributions to the CP violating decay of right
handed neutrino to Higgs and leptons, due to the extra Higgs and
extra gauge boson predicted in this model. We emphasized that
although these new contributions may be sizable, they have no
direct impact since they do not contain any strong CP violating
phase. Therefore, they contribute to lepton asymmetry via the
interference with one loop vertex and self-energy diagrams. In
this respect, a successful baryogenesis can be obtained in the
resonant leptogenesis scenario where the right handed neutrinos
are semi-degenerate in masses.


\end{document}